\documentclass[11pt, oneside]{article}   	
\usepackage{geometry}                		
\usepackage{pdfpages}
\usepackage{amsmath}
\usepackage{appendix}
\usepackage{cite}
\usepackage{graphicx}
\usepackage[hang]{subfigure}
\usepackage{bm}
\usepackage{float}
\usepackage{mathrsfs}
\usepackage{amssymb}

\newcommand{\bn}{\mathbf{n}}

\newcommand{\bx}{\mathbf{x}}
\newcommand{\md}{\mathrm{d}}

\makeatletter 
  \newcommand\figcaption{\def\@captype{figure}\caption} 
  \newcommand\tabcaption{\def\@captype{table}\caption} 
\makeatother

\begin{document}
\title{%
  {\huge \textit{Transition of defect patterns from 2D to 3D in liquid crystals}}
}

\author{Yang Qu, Ying Wei and Pingwen Zhang}
\date{}
\maketitle

\begin{abstract}
Defects arise when nematic liquid crystals are under topological constraints at the boundary. 
Recently the study of defects has drawn a lot of attention. In this paper, we 
investigate the relationship between two-dimensional defects and three-dimensional defects 
within nematic liquid crystals confined in a shell under the 
Landau-de Gennes model. We use a highly accurate spectral method to numerically solve the 
Landau-de Gennes model to get the detailed static structures of defects. Interestingly, the solution is radial-invariant 
when the thickness of the shell is sufficiently small. As the shell 
thickness increase, the solution undergo symmetry break to reconfigure the 
disclination lines. We study this three-dimensional reconfiguration of disclination 
lines in detail under different boundary conditions. 
We also discuss the topological charge of defects in two- and three-dimensional spaces 
within the tensor model. 
\end{abstract}

\section{Introduction}

\ \ \ \ \ Nematic liquid crystals (LCs) are composed of rigid rod-like molecules which tend to align in parallel 
with each other due to inter-molecular interaction. When the alignment of LC molecules is under topological 
constraints at the boundary, discontinuity in 
the alignment direction of LCs can form, which is known as defects. Typical defects include point defects
and disclination lines \cite{11}. The prediction of defect patterns is of great 
theoretical and practical interests, and remains to be a difficult problem. Because  defect patterns are not only 
affected by temperature and constraints, but also affected by the shape of the geometric regions bounding them. 

In recent years, lots of people have worked on nematic LCs confined on a sphere\cite{02,03}. 
It is assumed that the thickness of the LC-layer is negligible. As a result, the 
system can be simplified as a two-dimensional sphere. 
Four stable configurations are reported, namely the splay, tennis-ball, rectangle and cut-and-rotate splay \cite{03}.
The defect patterns of LC are also studied in three-dimensional spaces such as spherical droplets and outside solid balls.
Mkaddem and Gartland \cite{10} investigated systematically the defect patterns of nematic LCs confined in spherical droplet 
under radial anchoring condition. They obtained the radial hedgehog, ring disclination
and split-core solutions by assuming rotational symmetry around the z-axis.
Hu et al. \cite{01} obtained three configurations (single-core, double-core and split-core) in the spherical droplet 
under planar anchoring condition. When spherical colloids are dispersed in nematic LCs with radial 
anchoring condition on the surface, two configurations (Saturn-ring and dipole) are obtained in \cite{08}. After enforcing planar
anchoring condition on the surface, three stable configurations of boojum cores (single-core, double-core and split-core) are obtained in \cite{04}. In addition, Porenta et al. \cite{add1} investigated 
the effects of flexoelectricity and order electricity on defect cores in nematic droplets.											

Defects can be classified by their topological invariants. Lavrentovich \cite{22} elaborated 
the topological charge by homotopy group theory, which is called winding number in \cite{23}. In the case of two-dimensional 
(2D) space, winding number is an appropriate metric, and easy to be understood. But in the case of three-dimensional (3D) 
space, 
winding number and topological charge are different. We use the winding numbers on the normal plane 
of defect points to measure the disclination lines in this paper, and discuss the distinctions between the 
winding number and the topological charge of disclination lines in Sec. 3.2. 

In this paper, we study defect patterns in LCs confined in a shell. The relationship between defect patterns 
and the thickness of the shell, the temperature, and the boundary conditions is investigated systematically. 
We show that the final defect configuration is closely related with the geometry of the confined volume, 
the material properties and the topological constraints at the boundary. Interestingly, it is found that when 
the shell thickness is small, the defect patterns tend to be radial-invariant, meaning that the configuration of every 
layer are same up to a scaling constant in the shell. We call this kind of structure as 2D structure. When 
the shell becomes thicker, defects may undergo dramatic reconfigurations and do not hold the radial-invariant property. We call these defect patterns as 3D structure. In this paper, we study the reconfiguration between 
2D and 3D systematically, which has been studied by 
Vitelli and Nelson \cite{07} using a vector model. In Sec. 2, We first introduce the Landau-de Gennes model, 
and then discuss the boundary conditions after nondimensionalize the model. 
In Sec. 3, we obtain the free-energy minimum and visualize the main numerical results by numerically solving 
the model. The topological charge on 2D and 3D spaces are also discussed. Finally, some concluding remarks are made in Sec. 4.

\section{Landau-de Gennes Tensor Model and Numerical Method}

\subsection{Landau-de Gennes Model}
\ \ \ \ \ Landau-de Gennes model is a widely used model for nematic LCs systems, see more detail in \cite{11}, in which the alignment of 
nematic LCs is described by a symmetric, traceless $3 \times 3$ matrix, known as the $\mathbf{Q}$-tensor order 
parameter. Here the three eigenvalues of $\mathbf{Q}$ are written as $\lambda_1 , \lambda_2 , \lambda_3$. $\mathbf{Q}$ is isotropic when $\lambda_1=\lambda_2=\lambda_3 =0$. $\mathbf{Q}$ is uniaxial and can be expressed as $\mathbf{Q}=s(\bn\bn-\frac{\mathbf{I}}{3})$ when any two of the eigenvalues equal 
to each other. 
Otherwise $\mathbf{Q}$ is biaxial.

In LdG model, the energy functional takes the following form:
\begin{equation}
 F[\mathbf{Q}]=\int_{\Omega}f_b(\mathbf{Q})+f_e(\mathbf{Q})\md\bx.
\end{equation}
Here the bulk energy density is 
\begin{equation}
 f_b(\mathbf{Q})=\frac{A}{2}\mathrm{Tr}\mathbf{Q}^2-\frac{B}{3}\mathrm{Tr}\mathbf{Q}^3+\frac{C}{4}(\mathrm{Tr}\mathbf{Q}^2)^2,
\end{equation}
and the elastic energy density is 
\begin{equation}
 f_e(\mathbf{Q})=\frac{L}{2}\mathbf{Q}_{ij,k}\mathbf{Q}_{ij,k}.
\end{equation}
Where A, B and C are the temperature and material dependent constants, and $L$ 
is the elastic constant. Summation over repeated indices is implied and the comma indicates spatial derivative. 
In this paper, we study the defect patterns in spherical shell (the middle area between two concentric spherical surfaces) 
$\Omega=\{\bm{x}\ \lvert \ R_1\leq \|\bm{x}\|\leq R_2\}$. 
We nondimensionalize the model by defining the characteristic length 
$\xi_0=\sqrt{\frac{27CL_1}{B^2}}$, reduced temperature $t=\frac{27AC}{B^2}$ and region constants 
$\tilde{R}_i=\frac{R_i}{\xi_0}$, and rescaling the variables by 
$\tilde{\bm{x}}=\frac{\bm{x}}{\xi_0}, \bm{x}\in \Omega$,
$\tilde{\mathbf{Q}}=\sqrt{\frac{27C^2}{2B^2}}\mathbf{Q}$, 
$\tilde{F}=\xi_0^3\sqrt{\frac{27C^3}{4B^2L_1^3}}F$.
After dropping all the tildes in $\tilde{F}(\tilde{\mathbf{Q}})$, we obtain 
\begin{equation}
 F[\mathbf{Q}]=\int_{\Omega}\frac{t}{2}\mathrm{Tr}\mathbf{Q}^2-\sqrt{6}\mathrm{Tr}\mathbf{Q}^3+\frac{1}{2}(\mathrm{Tr}\mathbf{Q}^2)^2
 +\frac{1}{2}\mathbf{Q}_{ij,k}\mathbf{Q}_{ij,k}\md\bx.
 \label{eqn03}
\end{equation}
The integration is taken over the rescaled computational domain $\Omega$. 
After scaling, the eigenvalues of the scaled $\mathbf{Q}$ take values in 
$(\lambda_{min}, \lambda_{max})$, wi th $\lambda_{min}=-\frac{1}{3}\sqrt{\frac{27C^2}{2B^2}}$ and 
$\lambda_{max}=\frac{2}{3}\sqrt{\frac{27C^2}{2B^2}}$.

The reduced temperature $t$ appears only in the bulk energy term in the LdG model.
For $-\infty <t <1$, the nematic phase is energetically favored by the bulk 
energy. We use $\mathbf{Q}^+$ to represent the global energy minimum of the bulk energy with no boundary constraint, and it is given as follows
\begin{equation}
 \mathbf{Q}^+=s^+(\bn\bn-\frac{\mathbf{I}}{3}),
 \label{Q+}
\end{equation}
with 
\begin{equation}
 s^+=\sqrt{\frac{3}{2}}\cdot \frac{3+\sqrt{9-8t}}{4}.
  \label{eqn02}
\end{equation}
Under certain boundary condition, $\mathbf{Q}^+$ may no longer be the minimum 
of the total energy, and defects may arise. Defect patterns are the delicate balance between bulk energy, elastic energy and surface free-energy. The anchoring of molecules on spherical surface can be controlled by adding surface free-energy to the total energy. In the following, we consider two kinds of BCs, namely the radial anchoring condition and the planar anchoring condition.

\begin{enumerate}
 \item Radial anchoring condition
 
 First we consider the radial anchoring condition, in which the surface free-energy density is given by 
  \begin{equation}
  f_s=\omega(\mathbf{Q}(\bx)-\mathbf{Q}^+(\bx))^2,
 \end{equation}
 where $\mathbf{Q}^+$ and $s^+$ are given in Eq. (\ref{Q+}) and Eq. (\ref{eqn02}) at $\bx \in \partial \Omega$. $\omega$ is the constant which controls the relative strength of BC. The physical meaning of this BC is that 
 all the LC molecules align perpendicular to the sphere surface, with alignment strength equals to the $s^+$ that minimize the bulk energy. 
 
 \item Planar anchoring condition
 
 Next we consider the more complex planar anchoring condition
 \begin{equation}
  \mathbf{Q}\bx=\lambda_\nu\bx,
  \label{planar}
 \end{equation}
 where $\bx$ is the normal direction of the surface and $\lambda_{min}\leq \lambda_\nu < 0$ is a constant. $\lambda_\nu$ measures 
 the strength of compression imposed on the LC molecules at the boundary along the normal direction. When $\lambda_\nu=\lambda_{min}$, 
 all the molecules aligned in the plane perpendicular to $\bx$. We define the surface free-energy density as
 \begin{equation}
  f_s=\omega\|(\mathbf{Q}-\lambda_\nu\mathbf{I})\bx\|^2,\ \ \ \bx\in \partial \Omega.
 \end{equation}
 Where $\lambda_\nu$ can be different values and we assume $\lambda_\nu=-\frac{s^+}{3}$, in which $s^+$ is given in Eq. (\ref{eqn02}).
\end{enumerate}

A spherical shell is built by two concentric spheres, and they can have different 
topological constraints for LC molecules. For example, we can take a boundary 
condition combination that has radial anchoring condition for the inner surface and planar 
anchoring condition for the outer surface, or vice versa. 

\subsection{Numerical Method}
\ \ \ \ An equilibrium configuration corresponds to an energetically favored state, so our goal is to find 
$\mathbf{Q}(\bx)$ that minimize the total free-energy. Here we numerically solve the 
LdG model with the spectral method based on Zernike polynomial expansion \cite{19}. 
By using Zernike polynomial to expand $\mathbf{Q}$-tensor, 
we treat the expansion coefficients as variables and use BFGS algorithm \cite{20} to minimize the objective function. 
A function defined in a unit ball can be expanded in terms of Zernike polynomials, and the method is introduced in \cite{01}, so we do not go through the detail here. 

In order to expand $\mathbf{Q}$-tensor by using Zernike polynomials, we change the computational
regions to the regions inside unit ball. For the shell 
$\Omega=\{\mathbf{x}\ \lvert \ R_1 \leq \|\mathbf{x}\|\leq R_2\}$, we rescale the variables by 
$\hat{r}=\frac{r-R_1}{2(R_2-R_1)}+\frac{1}{2}$, 
$\hat{\theta}=\theta$, $\hat{\phi}=\phi$ ($(r, \theta, \phi)$ are spherical coordinates). 
As a result, the rescaled computational domain turns into $\hat{\Omega}=\{\mathbf{x}\ \lvert \ \frac{1}{2} \leq \|\mathbf{x}\|\leq 1\}$. This transformation applies only to the standard spherical shell, which means that $R_1 \neq 0$ and $R_2 \neq \infty$. 
More detail of this algorithm is described in the Appendix.

\section{Numerical Results}
\ \ \ \ To visualize the tensor field, we use ellipsoid to express $\mathbf{Q}$. Here three semi-principle axes of ellipsoid lie
in the eigenvectors of $\mathbf{Q}$ with lengths equal to the corresponding $\lambda_i+\frac{1}{3}$,  
in which $\lambda_i$ represents the eigenvalues of $\mathbf{Q}$ before rescaling. In this
representation, an isotropic $\mathbf{Q}$ is shown as a ball and a uniaxial $\mathbf{Q}$ with positive (negative) $s$ is 
prolate (oblate).

In order to describe the biaxial properties of the $\mathbf{Q}$-tensor, we follow \cite{10} and define
\begin{equation}
 \beta=1-6\frac{(tr\mathbf{Q}^3)^2}{(tr\mathbf{Q}^2)^3}.
\end{equation}
$\beta = 0$ represents pure uniaxial configuration while $\beta \neq 0$ corresponds 
to biaxial region. To visualize defects, following \cite{14}, we define
\begin{equation}
 c_l = \lambda_1 - \lambda_2,
 \label{eqn01}
\end{equation}
where $\lambda_1 \geq \lambda_2 \geq \lambda_3$. 
At defect point, $c_l = 0$, so we use the iso-surface of $c_l = \delta$ for a small positive 
constant $\delta$ to indicate where the defect is.

\subsection{The transform of defect pattern}
\ \ \ \ \ First we consider planar anchoring condition on both inner and outer surfaces of shell. 
In the limit that the thickness of the shell goes to zero, the system becomes the LC confined on sphere, 
which has been studied in \cite{02,03,17}. 
There are at least three defect configurations on the spherical surface reported \cite{02, 03}, namely the splay,
rectangle and tennis-ball. Cheng \cite{02} pointed out that splay and rectangle configurations are 
meta-stable in a 2D tensor model. When the thickness of the shell is small but is not zero, 
splay and rectangle configurations are always unstable. During the optimization 
process, the local derivatives near these configurations can be very small, but they will eventually lose stability 
and switch to the tennis-ball configuration for a sufficiently long iteration. In this case, a very thin spherical shell 
is not entirely equivalent to a spherical surface. It is unrealistic to treat molecules on the sphere have zero thickness. 

After combining the above inner and outer BC, the defect structures in the shell become 2D defect namely 
$\mathbf{Q}$ has similar, even uniform distribution on arbitrary spherical surface of shell when a shell is thin enough. Fig. 
\ref{pic16} illustrates three 2D configurations for different temperature $t$ and outer radius $R_2$. Fig. \ref{pic16} 
(a) is a tennis-ball solution in which four $+1/2$ short disclination lines are inscribed 
in the shell. The disclination lines start appear from the inner surface and extend to the outer surface, and the lines form a regular tetrahedron on each layer between two surfaces. In the case of thin shell, Fig. \ref{pic16} (b) and (c) give two other 2D configurations which only exist at high temperature. Fig. \ref{pic16} (b) shows a 2D oblate solution which is similar to the radial hedgehog solution, in which 
$\mathbf{Q}$ is uniaxial everywhere and satisfies radial symmetry. The director vector $\bn$ align with the direction of radius on each point in Fig. \ref{pic16} (b), which gives no defect in this configuration. 
Fig. \ref{pic16} (c) shows a 2D single-single core solution which composed by two $+1$ disclination lines 
(an extension of two single-core configurations) in the shell. It is an axial symmetric solution that has oblate 
$\mathbf{Q}$-tensor along the symmetric axis. In addition, we find that the configuration shown in Fig. \ref{pic16} (c) disappears and converts to the configuration shown in (b) as the shell becomes thinner.
\\[\intextsep]
  \begin{minipage}{\textwidth} 
    \centering 
    \clearpage
    \includegraphics[width=0.95\textwidth]{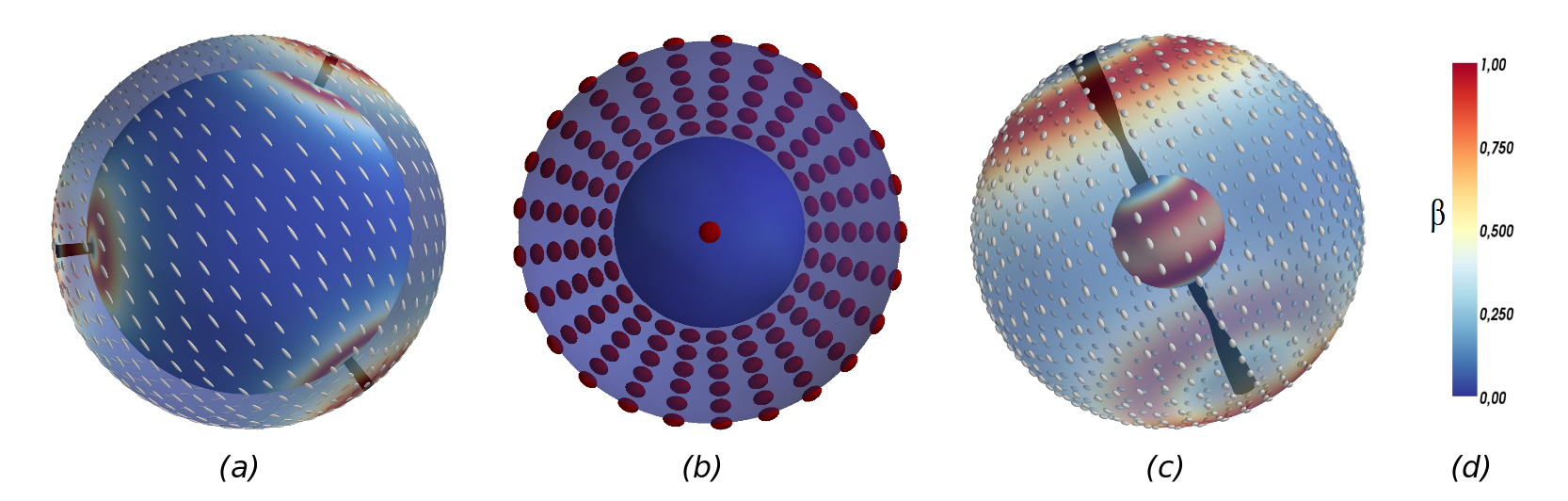}
    \figcaption{The 2D solutions for the planar anchoring condition on inner and outer spheres for 
    different $t$ and $R_2$ (a: $t = -5, R_2=1.2$; b:  $t = 0.8, R_2=2$; c:  $t = 0.8, R_2=3.34$) with $R_1=1$. 
    $\beta$ is shown in color with red corresponds to biaxial and blue uniaxial. Ellipsoids represent the $\mathbf{Q}$-tensor. 
    (a): Three-dimensional view of tennis-ball solution; (b): A sliced view of the uniaxial solution for the thin shell; (c): 2D 
    single-single core solution. The black pipes inside the shell are the iso-surface of $c_l$ with values equal 
    to: (a) $c_l=0.04$, (c) $c_l = 0.002$.}
    \label{pic16}
  \end{minipage} 
\\[\intextsep] 

As the thickness of the shell 
increases, many 3D configurations appear as shown in Fig. \ref{pic10}. Fig. \ref{pic10} (a) shows an 
axially symmetric solution (also was mentioned in \cite{07}) which contains four segments of $+1$ disclination lines 
on the two poles of the surfaces. For all these four disclination lines in Fig. \ref{pic10} (a), an endpoint of each line is isotropic and buried inside the shell while the another endpoint stays on the sphere. Around each defect point on disclination lines, 
there is a ring of biaxial region. The state in Fig. \ref{pic10} (a) only exists at high temperature like the 
two 2D configurations in Fig. \ref{pic16}. As the thickness of the shell increases, the stable 
solution converts from Fig. \ref{pic16} (b) to Fig. 
\ref{pic16} (c), and then transform to Fig. \ref{pic10} (a). 
Fig. \ref{pic10} (b-d) show inner double-core plus outer double-core, inner double-core plus outer split-core and 
inner split-core plus outer split-core configurations respectively. 
A common feature of these structures shown in Fig. \ref{pic10} (b-d) is that each structure has four disclination lines in which two are parallel to each other on one side, and two are perpendicular to each other on opposite sides. Fig. \ref{pic10} (e-h) 
show $\beta$ and $\mathbf{Q}$ inside the shells. It is easy to see that all the configurations above are the combinations of single-core, double-core and split-core. In our results, the defect patterns derived by the outer sphere are similar to the configurations in the spherical droplet \cite{01}. And the defect patterns caused by the inner sphere  are similar to the configurations outside of the solid ball \cite{08, 04}. 
\\[\intextsep] 
  \begin{minipage}{\textwidth} 
    \centering 
    \clearpage
    \includegraphics[width=0.9\textwidth]{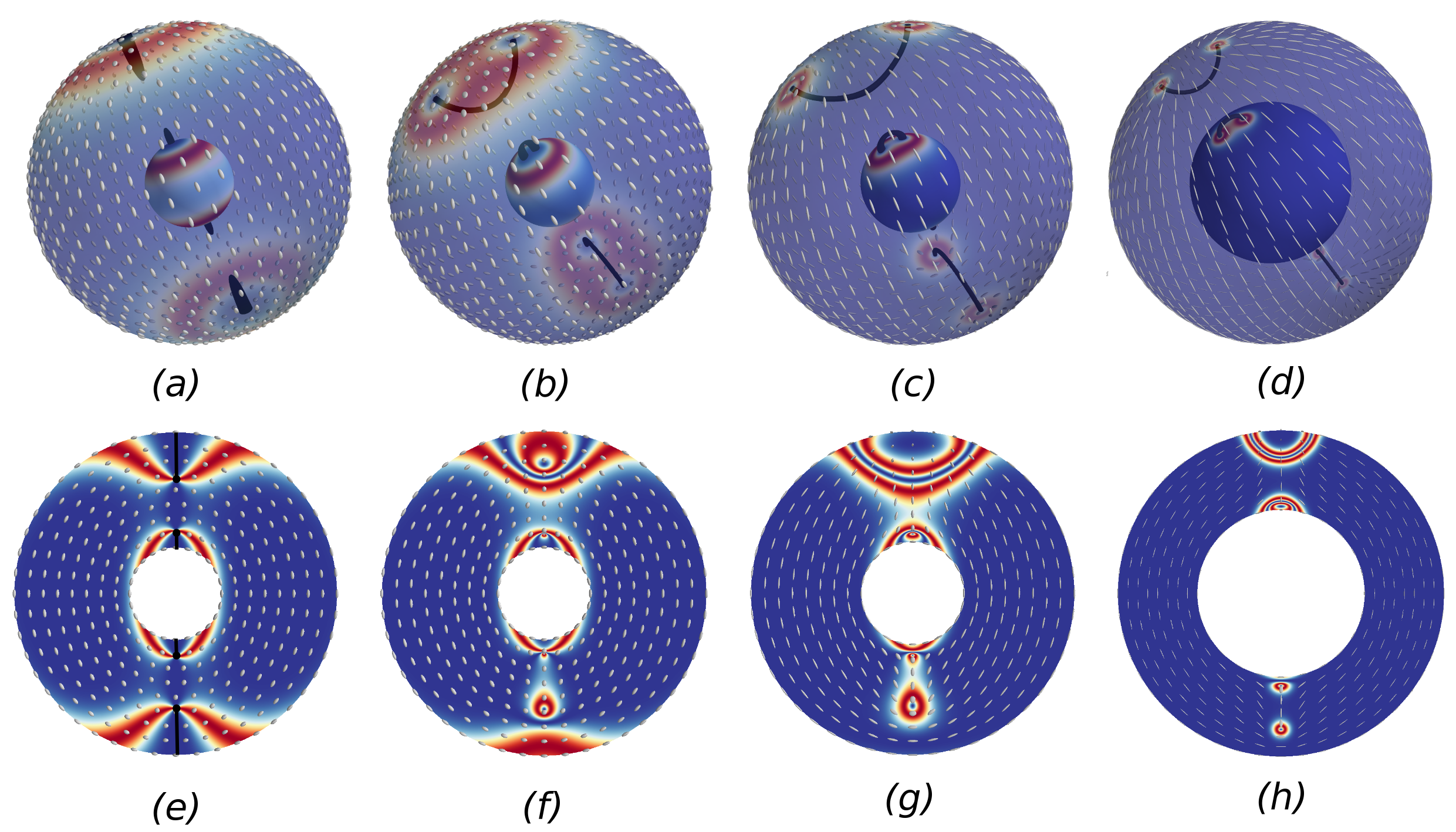}
    \figcaption{Four stable solutions for the planar anchoring condition on inner and outer spheres for different 
    $t$, $R_1$ and $R_2$ (a and e: $t = 0.5, R_1=1, R_2=3.5$; b and f:  $t = 0.1, R_1=1, R_2=3.5$; c and g:  $t = -4, R_1=1, 
    R_2=3.15$; d and h:  $t = -8, R_1=4, R_2=7.8$). The ellipsoids represent the 
    $\mathbf{Q}$-tensor. Color corresponds to $\beta$, ranging from 0 (blue) to 1 (red). The black pipes inside 
    the shell in (a-d) are the iso-surface of $c_l$ , with values equal to (a): $c_l = 0.002$; (b): $c_l = 0.012$; 
    (c): $c_l = 0.05$; (d): $c_l = 0.15$.(a-d) Three-dimensional view. 
    (e-h) Sliced view showing the inside of the LC shell. The cutting plane in (e) pass through the symmetric axis. And the 
    cutting plane in (f-h) is determined by the sphere center and a pair of defect points on the surface. The thick black lines 
    in (e) represent four segments of $+1$ disclination lines.
    }
    \label{pic10}
  \end{minipage} 
\\[\intextsep] 

When the shell is thin, the defect has 2D structures, and when shell is thick, the defect 
has 3D structures. We aim to study how the 2D structures lose their stability and change to 
the 3D structures as the thickness of the shell increases, we increase $R_2$ slowly and observe how 
the defect pattern change when $R_1$ is fixed. We find that for the low temperature, the four disclination lines 
in Fig. \ref{pic16} (a) will bend into the shapes in Fig. \ref{pic31} (a) as $R_2$ increases. Because of this special defect structure, the four defect points on each layer no longer keep 
a regular tetrahedron shape, change to a tetrahedron which has four congruent isosceles triangles. 
Consider the fixed temperature $t=-5$, and if $R_2>R^*$ where $R^*$ is the transition point, then the 
configuration change from Fig. \ref{pic31} (a) to Fig. \ref{pic31} (b).
During this process, the two pairs of disclination lines do not touch each other. At high 
temperature, Fig. \ref{pic32} (a) shows that the tennis-ball structure changes to the structure shown in (b) 
in which a pair of disclination lines touch each other on an intermediate point as $R_2$ increases. 
Increasing the thickness, the disclination lines shown in (b) split into four hemicycle disclination lines, 
and the configuration change to  double-split core as given in Fig. \ref{pic32} (c).
Different from the discontinuous transition from 2D to 3D, it is continuous in the high temperature.
\\[\intextsep] 
  \begin{minipage}{\textwidth} 
    \centering 
    \clearpage
    \includegraphics[width=0.75\textwidth]{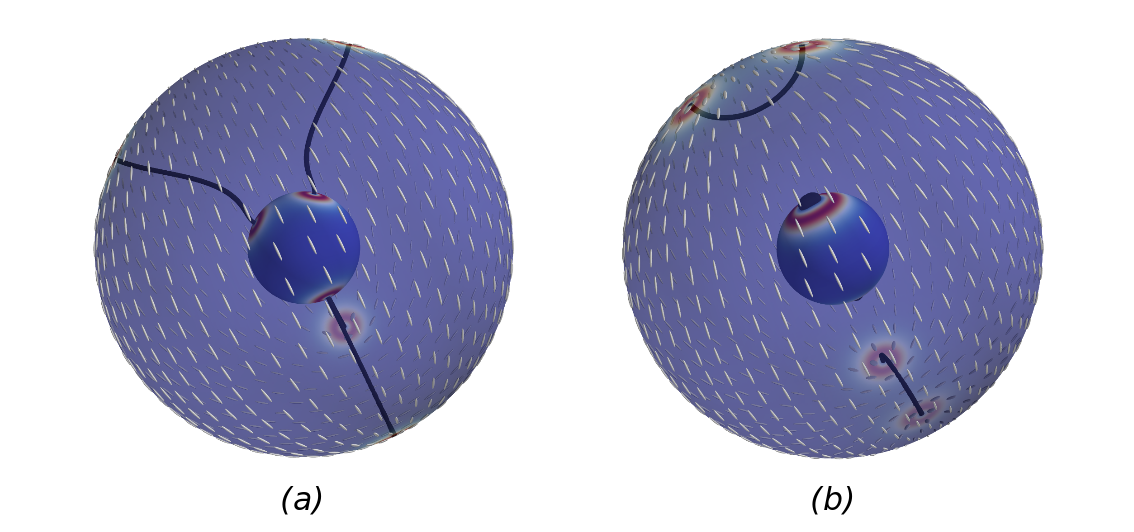}
    \figcaption{Two stable solutions under temperature $t = - 5, R_1=1$ with planar anchoring condition on inner and outer 
    surfaces of shell. Outer radii $R_2$ equal to (a: $R_2 = 3.6$; b: $R_2 = 3.61$). The iso-surface of $c_l$ has the value 
    equals to $c_l$ = 0.05 in (a-b).} 
    \label{pic31} 
  \end{minipage} 
\\[\intextsep]
\\[\intextsep] 
  \begin{minipage}{\textwidth} 
    \centering 
    \clearpage
    \includegraphics[width=0.9\textwidth]{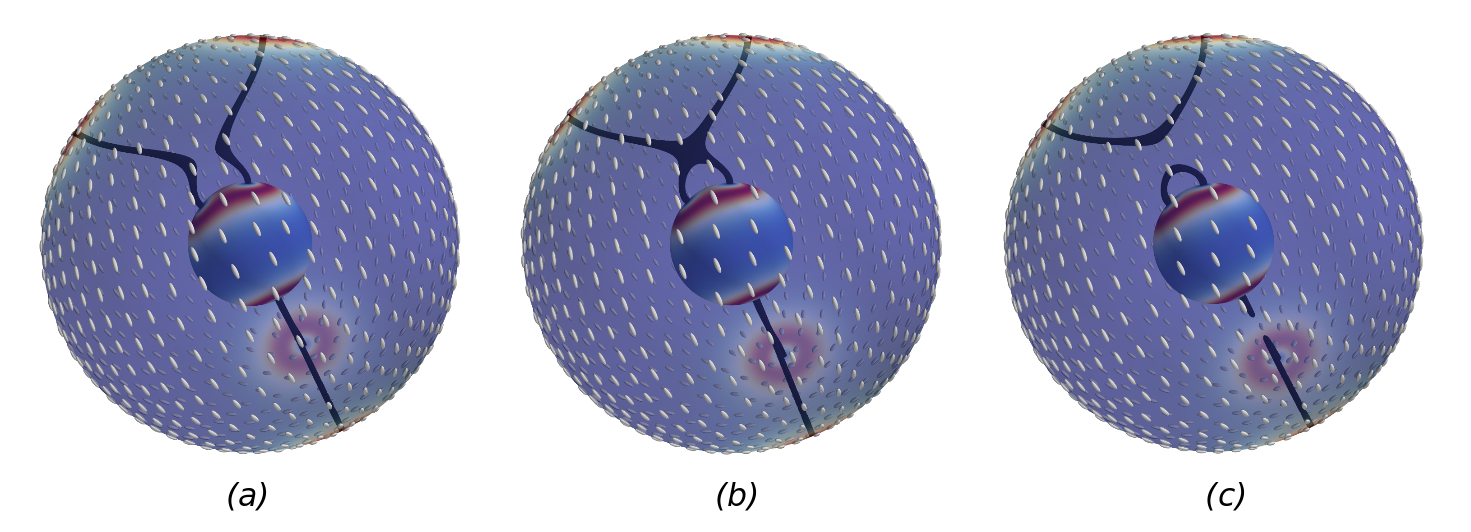}
    \figcaption{Three stable solutions under temperature $t = - 0.4, R_1=1$ with planar anchoring condition on inner and outer 
    surfaces. Outer radii $R_2$ equal to (a: $R_2 = 3.25$; b: $R_2 = 3.29$; c: $R_2 = 3.34$). The iso-surface of $c_l$ has 
    the value equals to $c_l$ = 0.02 in (a-c).} 
    \label{pic32} 
  \end{minipage} 
\\[\intextsep]

Here we call the defect pattern shown in Fig. \ref{pic31} (a) 2D configuration, and call the defect pattern shown in Fig. \ref{pic32} (b) an intermediate state between 2D and 3D configurations. We compute the phase diagram shown in Fig. \ref{pic12} with $R_1$ equals to 1, and confirm that there is no transition region when the configuration changes from 2D to 3D under low temperature assumption. The left side of dotted-dashed line keeps only 2D configuration, and the right side of dashed line keeps only 3D configuration in the phase diagram. 
The two configurations exist in the region between the dotted-dashed and dashed lines, in which the 2D (3D) configuration has less energy than the 3D (2D) configuration in the region between the dotted-dashed (dashed) line and the solid line. We define the temperature below $t^*$ given in Fig. \ref{pic12} is low temperature, and the temperature above $t^*$ is high temperature. From the phase diagram, it also approve that the transition between 2D and 3D configurations is discontinuous when $t<t^*$, and it's continuous when $t>t^*$ as we observed in numerical results.
\\[\intextsep] 
  \begin{minipage}{\textwidth} 
    \centering 
    \clearpage
    \includegraphics[width=0.55\textwidth]{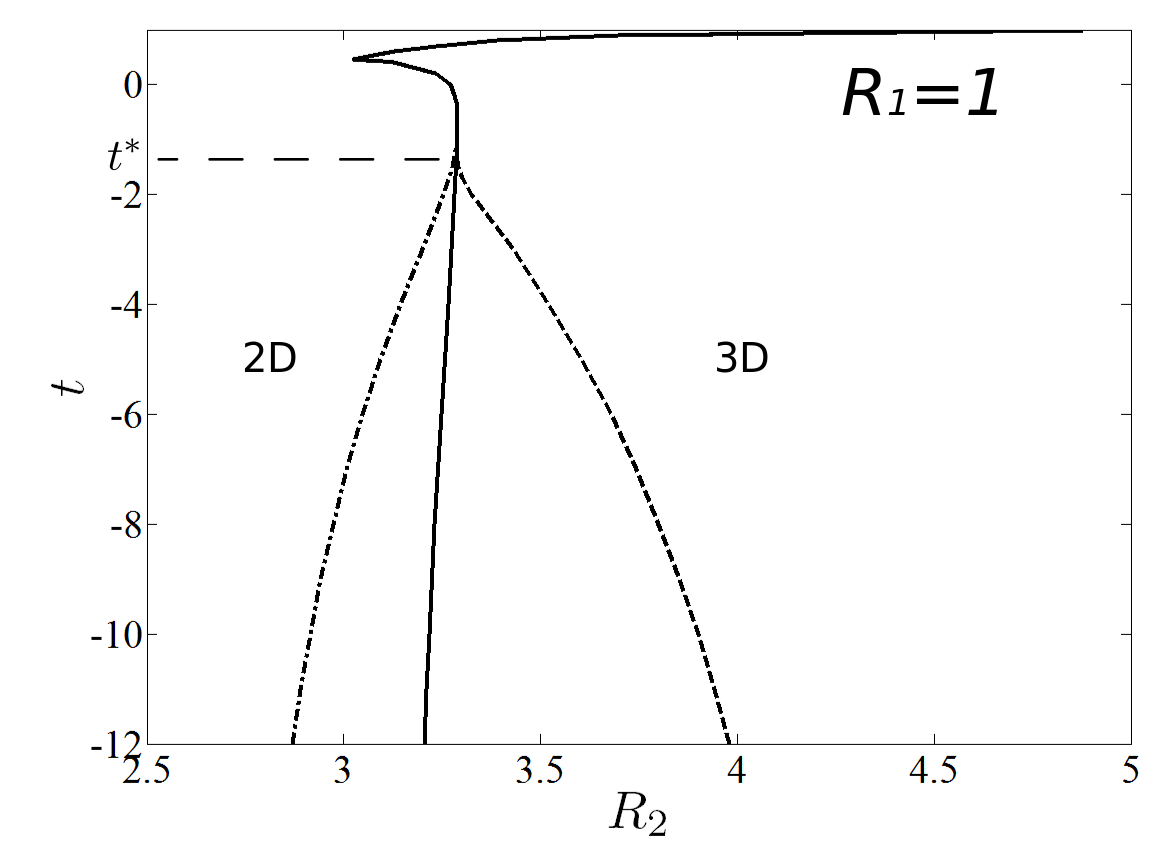}
    \figcaption{Phase diagram for the planar anchoring condition on both inner and outer spheres. 
    Left to the dotted-dashed line, 3D configuration doesn't exist. Right to the dashed line, 2D configuration doesn't exist. 
    The solid line in the middle is total free-energy dividing line. 2D solutions 
    are  global free-energy minimizer on the left side of this line. on the other side, 3D solutions are global minimizer.} 
    \label{pic12} 
  \end{minipage} 
\\[\intextsep] 

For the spherical shell with radial anchoring conditions on both the inner and outer spheres, we obtain the 
radial-hedgehog solution in which $\mathbf{Q}$ is prolate everywhere in the shell and satisfies the radial symmetry. Majumdar et al. studied the stability of the radial-hedgehog solution on 3D spherical shell in \cite{25}. They proved that the radial-hedgehog solution is the unique global Landau-de Gennes energy minimum for a sufficiently narrow spherical shell, for all temperatures 
below the nematic supercooling temperature. 

Below we consider radial anchoring condition on inner sphere and planar anchoring condition on outer sphere.
When the thickness of shell is large enough, the structures we get are the combinations of single-core, double-core, split-core and Saturn-ring. But as the thickness decrease, these structures change also. Consider the fixed temperature $t=-3$, the change process is shown in Fig. \ref{pic15}. Fig. \ref{pic15} (a) shows a configuration which has one $-1/2$ disclination ring around  the inner sphere and two hemicycle disclination lines on the outer sphere. Obviously this structure is a combination of split-core and Saturn-ring. When we 
decrease the outer sphere radius in Fig. \ref{pic15} (a) to $2.4$, we obtain the solution in Fig. \ref{pic15} (b) which has some distortion. Because of the interaction between two kinds of disclination lines, the disclination ring in the shell curve upward on one diameter direction, and curve downward on the other vertical direction, like a saddle shape. This shows that the disclination lines feel interaction to each other when the thickness of shell is small. Decreasing the thickness, we can get a uniaxial solution as shown in Fig. \ref{pic15} (c).  $\mathbf{Q}$ in this configuration is uniaxial everywhere and satisfies the radial symmetry. $\mathbf{Q}$ changes from prolate to oblate and $s$ changes from $s^+$ to $-s^+/2$ continuously as radius increases. This configuration can be regard as a 2D configuration because the director vector $\bn$ has 
uniform distribution on every layer of spherical shell. 
\\[\intextsep] 
  \begin{minipage}{\textwidth} 
    \centering 
    \clearpage
    \includegraphics[width=0.9\textwidth]{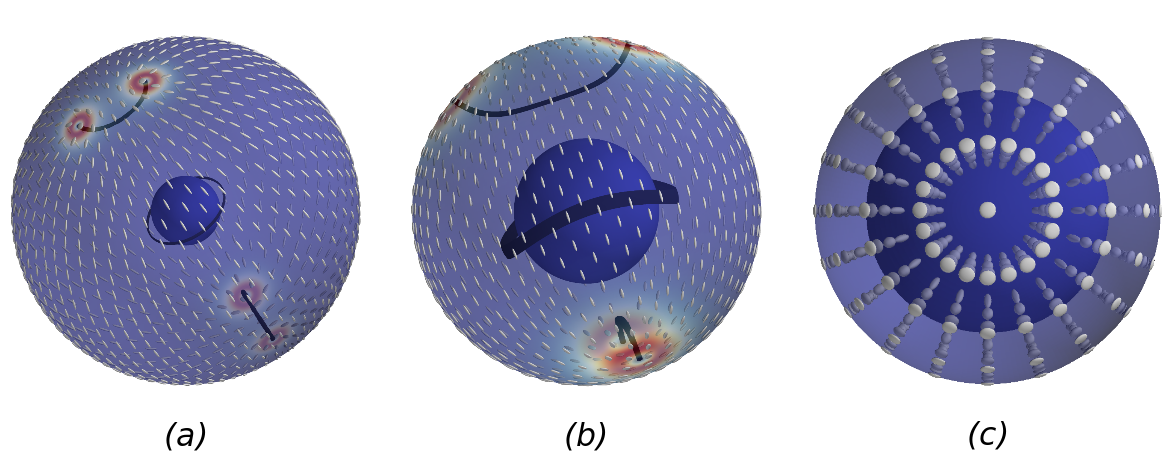}
    \figcaption{Three stable solutions under temperature $t = -3, R_1=1$ with radial anchoring condition on inner surface 
    and planar anchoring condition on outer surface. Outer radii $R_2$ equal to (a: $R_2 = 5$; b: $R_2 = 2.4$; c: $R_2 = 1.4$). 
    The iso-surface of $c_l$ has the values equal to: (a) $c_l=0.05$, (b) $c_l = 0.03$.}
    \label{pic15}
  \end{minipage} 
\\[\intextsep]

\subsection{The discussion of topological charge}

\ \ \ \ In this subsection we mainly discuss the topological charge of defects in 2D and 3D spaces.
For the case of 2D space, winding number is an appropriate metric, because only
point defects exist. But for the case of 3D space, winding number and topological
charge are different. Fig. \ref{pic37} shows two part views of 
split-core on the inner surface of the shell. The black short lines in figures represent the eigenvector corresponding to the largest eigenvalue. Fig. \ref{pic37} (a) displays the distribution of LC molecules near the defect points on the inner sphere. 
It's easy to see that the winding numbers of two defect points on the surface are $+1/2$.
Fig. \ref{pic37} (b) displays the distribution of LC molecules on the normal plane of disclination line's midpoint, which has the winding number equals to $-1/2$. 
We know that the disclination lines in 3D space can be classified according to their homotopy, in which topological charge only has $0$ and $1/2$ two categories. The above two kinds of defect points in split-core both are in the $1/2$ category. So winding number as an intuitive concept no longer match the topological charge in 3D space. Winding number is not good to describe the global properties of disclination lines. 
Hu et al. \cite{01} investigated nematic LCs on 2D disk. They obtained the $\pm k/2$-disclination lines for $k>1$, and observed that $\pm k/2$-disclination lines quantize to give k $\pm 1/2$-disclination lines under low temperature. 
This shows that the simple classification determined by topological charge can not depict the variety winding numbers of 
disclination lines clearly. This give us a reason to think how to effectively define the winding number in 3D space 
to crisply distinguish various kinds of disclination lines. 
\\[\intextsep] 
  \begin{minipage}{\textwidth} 
    \centering 
    \clearpage
    \includegraphics[width=0.9\textwidth]{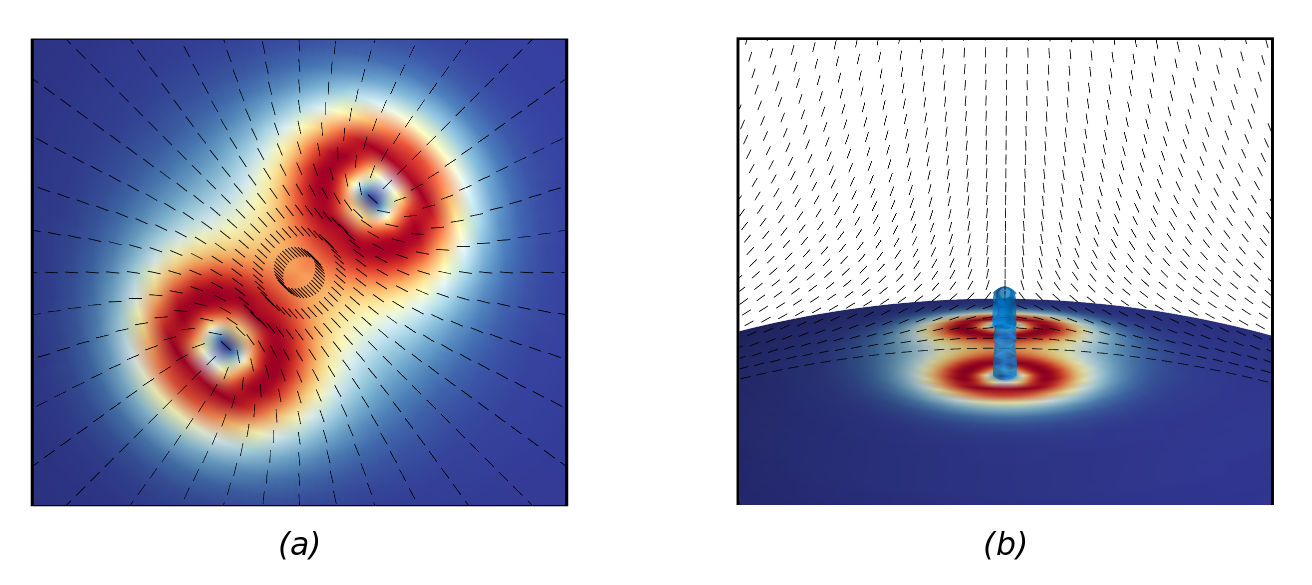}
    \figcaption{Part views of split-core configuration on the inner surface of shell. (a) displays the distribution of LC 
     molecules near the defect points on the spherical surface. (b) displays the distribution of LC molecules on the 
     normal plane of the disclination line's midpoint. The black short lines in figures represent the eigenvector corresponding to the largest eigenvalue. The pipe in (b) is the iso-surface of $c_l=0.08$.}
    \label{pic37}
  \end{minipage} 
\\[\intextsep]

\section{Discussion and Conclusion}

\ \ \ \ In this paper, we investigate defect patterns in three-dimensional spherical shell with LdG model. By using the 
spectral method, we can calculate many interesting configurations under different parameters. Based on 
the observations made from these results, we are ready to summarize some general rules. 
First we try to understand the relationship between 2D spherical surface and 3D spherical shell, we find 
that the structures exist on the spherical surface are not necessarily exist in the thin spherical shell. 
It means that some configurations only occur when the thickness of LC molecules 
layer equals to zero, which is unrealistic. So the difference between spherical surface and thin spherical shell 
is essential. Then we study the transform between 2D configurations and 3D configurations in spherical shells. 
When the thickness of shell is large enough, the defect patterns are 3D configuration, which are the combinations 
of defect patterns derived by two boundary surfaces. But when the thickness of shell is thin enough,2D configurations 
are energetically favored because of the lack of radial space. For the shell with planar boundary condition on both 
inner and outer surfaces, the transform between 2D configurations and 3D configurations can be divided into continuous 
and discontinuous by temperature.

\ \ 

\textbf{Acknowledgement} Dr. Yucheng Hu provides many useful suggestions. P. Zhang is partly supported by National Natural Science Foundation of China (Grant No. 11421101 and No. 11421110001).

\section{Appendix}

\subsection{Numerical Approach}
\ \ \ \ A stable liquid crystal phase corresponds to a minimum energy state, so our goal is to find the $\mathbf{Q}$-tensor distribution minimizing 
the total free-energy functional. Here we use spectral method to numerically solves the Landau-de Gennes model. Based on Zernike polynomial 
expansion, the free-energy can be expressed as a function of expansion coefficients. Then we can use optimization methods such as BFGS algorithm 
to search for local energy minimum. 
A function defined in a unit ball can be expanded in terms of Zernike polynomials, which are introduced detailedly in \cite{01}.

In order to expand $\mathbf{Q}$-tensor by using Zernike polynomials, we change the computational regions to the regions inside unit ball. For the shell 
$\Omega=\{\mathbf{x}\ \lvert \ R_1 \leq \|\mathbf{x}\|\leq R_2\}$, we rescale the variables by $\hat{r}=\frac{r-R_1}{2(R_2-R_1)}+\frac{1}{2}$, 
$\hat{\theta}=\theta$, $\hat{\phi}=\phi$. The corresponding coordinate transformation under the Cartesian coordinate system are
\begin{eqnarray}
 \begin{array}{ccc}
   \hat{x}=\frac{x}{2(R_2-R_1)}+\frac{R_2-2R_1}{2(R_2-R_1)}\frac{x}{r},\\
   \hat{y}=\frac{y}{2(R_2-R_1)}+\frac{R_2-2R_1}{2(R_2-R_1)}\frac{y}{r},   \\ 
   \hat{z}=\frac{z}{2(R_2-R_1)}+\frac{R_2-2R_1}{2(R_2-R_1)}\frac{z}{r}.   \\
   \end{array}
\end{eqnarray}
The rescaled computational domain is $\hat{\Omega}=\{\mathbf{x}\ \lvert \ \frac{1}{2} \leq \|\mathbf{x}\|\leq 1\}$. The free-energy functional becomes
\begin{equation}
  F[\mathbf{Q}]=\int_{0}^{2\pi}\int_{0}^{\pi}\int_{\frac{1}{2}}^{1}(\frac{t}{2}\mathrm{Tr}\mathbf{Q}^2-\sqrt{6}\mathrm{Tr}\mathbf{Q}^3+\frac{1}{2}(\mathrm{Tr}\mathbf{Q}^2)^2
 +\frac{1}{2R^2}\mathbf{Q}_{ij,k}\mathbf{Q}_{ij,k})E(\hat{r})\sin\hat{\theta} \md\hat{r} \md\hat{\theta} \md\hat{\phi}.
\end{equation}
Where $E(\hat{r})=A\hat{r}^2+B\hat{r}+C$, 
\begin{equation}
 \begin{array}{l}
  A = 8(R_2-R_1)^3,\\
  B = 8(R_2-R_1)^2(2R_1-R_2),\\
  C = 2(R_2-R_1)(2R_1-R_2)^2.
 \end{array}
\end{equation}
Where $\mathbf{Q}_{ij,k}$ still expresses the partial derivatives of $\mathbf{Q}$ respect to original coordinates $x, y, z$. Using $(q_x, q_y, q_z)$ 
express the partial derivatives of components of $\mathbf{Q}$ respect to $x, y, z$, the partial derivatives can be calculated by the following formula
\begin{equation}
\left(\begin{array}{c}
q_{x} \\
q_y \\
q_z \\
\end{array}
\right)=
\left(\begin{array}{ccc}
\frac{\partial \hat{x}}{\partial x} & \frac{\partial \hat{y}}{\partial x} & \frac{\partial \hat{z}}{\partial x}\\
\frac{\partial \hat{x}}{\partial y} & \frac{\partial \hat{y}}{\partial y} & \frac{\partial \hat{z}}{\partial y}\\
\frac{\partial \hat{x}}{\partial z} & \frac{\partial \hat{y}}{\partial z} & \frac{\partial \hat{z}}{\partial z}\\
\end{array}
\right)
\left(\begin{array}{c}
q_{\hat{x}} \\
q_{\hat{y}} \\
q_{\hat{z}} \\
\end{array}
\right)=
J\left(\begin{array}{c}
q_{\hat{x}} \\
q_{\hat{y}} \\
q_{\hat{z}} \\
\end{array}
\right).
\label{qxyz}
\end{equation}
The transformation matrix is as follows
\begin{equation}
J=
\left(\begin{array}{ccc}
\frac{1}{2(R_2-R_1)}+\frac{R_2-2R_1}{2(R_2-R_1)}\frac{r^2-x^2}{r^3} & -\frac{R_2-2R_1}{2(R_2-R_1)}\frac{xy}{r^3} & -\frac{R_2-2R_1}{2(R_2-R_1)}\frac{xz}{r^3}\\
-\frac{R_2-2R_1}{2(R_2-R_1)}\frac{xy}{r^3} & \frac{1}{2(R_2-R_1)}+\frac{R_2-2R_1}{2(R_2-R_1)}\frac{r^2-y^2}{r^3} & -\frac{R_2-2R_1}{2(R_2-R_1)}\frac{yz}{r^3}\\
-\frac{R_2-2R_1}{2(R_2-R_1)}\frac{xz}{r^3} & -\frac{R_2-2R_1}{2(R_2-R_1)}\frac{yz}{r^3} &\frac{1}{2(R_2-R_1)}+\frac{R_2-2R_1}{2(R_2-R_1)}\frac{r^2-z^2}{r^3}\\
\end{array}
\right).\nonumber
\end{equation}

Because $\mathbf{Q}$ is a traceless symmetrical matrix, so it can be expressed as the following form:
\begin{eqnarray}
 \mathbf{Q}=\left(\begin{array}{ccc}
                      q_1 & q_2 &q_3\\
                      q_2 & q_4 &q_5\\ 
                      q_3 & q_5 &-q_1-q_4\\
                     \end{array}
\right).
\end{eqnarray}
Here we expand $q_i, i=1,\cdots , 5$ in Zernike polynomials as follows
\begin{equation}
 q_i(r,\theta, \phi)=\sum_{m=1-M}^{M-1}\sum_{l=\lvert m \rvert}^{L-1}\sum_{n=l}^{N-1}
 A_{nlm}^{(i)}Z_{nlm}(r,\theta,\phi).
 \label{zernike}
\end{equation}
Where $N\geq L\geq M \geq 0$. So given $A_{nlm}^{(i)}$, we can calculate the free-energy and partial derivatives of 
free-energy function respect to $A_{nlm}^{(i)}$ by numerical integration.


\begin{thebibliography}{10}
\bibitem{01} { Y. C. Hu, Y. Qu, and P. W. Zhang}, {\em On the Disclination Lines of Nematic Liquid
Crystals}, arXiv:14086191, (2014).

\bibitem{02} { H. Cheng and P. W. Zhang}, {\em A tensor model for liquid crystals on a spherical
surface}, Sci. China Math., 56 (2013), pp. 2549-€"2559.

\bibitem{03} { W. -Y. Zhang, Y. Jiang, and J. Z. Y. Chen}, {\em Onsager model for the structure
of rigid rods confined on a spherical surface}, Phys. Rev. Lett., 108 (2012), p. 057801.

\bibitem{22} { O. D. Lavrentovich}, {\em Defects in Liquid Crystals: Surface and Interfacial Anchoring Effects}, 
in Patterns of Symmetry Breaking, Springer, 127 (2003), pp. 161-195.

\bibitem{23} { T. C. Lubensky, David Pettey, Nathan Currier, and Holger Stark}, {\em Topological 
defects and interactions in nematic emulsions}, Phys. Rev. E, 57 (1998), p. 610.

\bibitem{04} { M. Tasinkevych, N. Silvestre, and M. M. Telo da Gama}, {\em Liquid crystal
boojum-colloids}, New J. Phys., 14 (2012), p. 073030.

\bibitem{05} { D. R. Nelson}, {\em Toward a tetravalent chemistry of colloids}, Nano Lett., 2 (2002),
pp. 1125-1129.

\bibitem{06} { T. Lopez-Leon, V. Koning, V. Vitelli, K. B. S. Davaiah, and A. Fernandez-Nieves}, {\em Frustrated
nematic order in spherical geometries}, Nature Physics, 7 (2011), p. 391.

\bibitem{07} { V. Vitelli and D. R. Nelson}, {\em Nematic texures in spherical shells}, 
Phys. Rev. E, 74 (2006), p. 021711.

\bibitem{08} { M. Ravnik and S. Žumer}, {\em Landau-de Gennes modelling of nematic liquid crystal 
colloids}, Liq. Cryst., 36 (2009), p. 1201.

\bibitem{18} { P. Poulin and D. A. Weitz}, {\em Inverted and multiple nematic emulsions}, Phys. Rev. E, 57 (1998), p. 626.

\bibitem{17} { S. Kralj, R. Rosso, and E. G. Virga}, {\em Curvature control of valence on nematic shells}, Soft Matter, 7 (2011), pp. 670-683.

\bibitem{09} { A. Majumdar}, {\em Equilibrium order parameters of nematic liquid crystals in the
Landau-de Gennes theory}, Eur. J. Appl. Math, 21 (2010), pp. 181-203.

\bibitem{10} { S. Mkaddem and E. Gartland Jr}, {\em Fine structure of defects in radial nematic
droplets}, Phys. Rev. E, 62 (2000), p. 6694.

\bibitem{add1} { T. Porenta, M. Ravnik,  and S. Zumer}, {\em Effect of flexoelectricity and order electricity on defect cores in nematic droplets}, Soft Matter, 7 (2011), p. 132.

\bibitem{11} { P. G. de Gennes and J. Prost}, {\em The physics of liquid crystals}, 
Oxford University Press, Oxford, second ed., 1995.

\bibitem{12} { F. C. Frank}, {\em On the theory of liquid crystals}, Discussions of Faraday Society, 
25 (1958), pp. 19-28.

\bibitem{13} { F. H. Lin and C. Liu}, {\em Static and dynamic theories of liquid crystals}, 
J. Partial Differ. Equ., 14 (2001), pp. 289-330.

\bibitem{14} { A. C. Callan-Jones, R. A. Pelcovits, V. A. Slavin, S. Zhang, D. H.
Laidlaw, and G. B. Loriot}, {\em Simulation and visualization of topological defects in nematic liquid crystals},
Phys. Rev. E, 74 (2006), p. 061701.

\bibitem{15} { D. S. Miller, X. Wang, and N. L. Abbott
}, {\em Design of functional materials based on liquid crystalline droplets},
Chem. Mater., 26 (2013), pp. 496-506.

\bibitem{16} { S. Kralj, E. G. Virga, and S. Zumer}, {\em Biaxial torus around nematic point defects},
Phys. Rev. E, 60 (1999), pp. 1858-1866.

\bibitem{19} { F. Zernike}, {\em Diffraction theory of the cut procedure and its improved form, the
phase contrast method}, Physica, 1 (1934), pp. 689-704.

\bibitem{20} { M. Avriel}, {\em Nonlinear programming: analysis and methods}, Courier Dover Publications, (2003).

\bibitem{21} { T. Lopez-Leon and A. Fernandez-Nieves}, {\em Drops and shells of liquid crystal}, Colloid Polym
Sci., 289 (2011), pp. 345–359.

\bibitem{25} { A. Majumdar, G. Canevari, and M. Ramaswamy}, {\em Radial symmetry on three-dimensional shells in the Landau-de Gennes theory}, arXiv:14090143, (2014).
\end{thebibliography}
\end{document}